\documentclass[12pt,a4paper]{article}
\usepackage{graphics}
\newcommand{\be}{\begin{equation}}
\newcommand{\ee}{\end{equation}}
\newcommand{\bea}{\begin{eqnarray}}
\newcommand{\eea}{\end{eqnarray}}

\newcommand{\href}[2]{#2}
\newcommand{\eprint}[1]{{\tt #1}}

\begin{document}



\title{The Effect of Weak Interactions on the Ultra-Relativistic
Bose-Einstein \\Condensation Temperature\thanks{{\tt
Imperial/TP/00-1/6}, \eprint{hep-ph/0011286}, 11th September
2001}}

\author{D.~J.~Bedingham\thanks{email:
\href{mailto:DJ.Bedingham@ic.ac.uk}{{\tt
 DJ.Bedingham@ic.ac.uk}}}
 and
 T.~S.~Evans\thanks{email: \href{mailto:T.Evans@ic.ac.uk}{{\tt
 T.Evans@ic.ac.uk}}}
 \\
 \href{http://theory.ic.ac.uk/}{Theoretical Physics},
 The Blackett Laboratory,
 Imperial College,\\
 Prince Consort Road,
 London, SW7 2BW, U.K.}


\maketitle

\begin{abstract}

We calculate the ultra-relativistic Bose-Einstein condensation
temperature of a complex scalar field with weak
$\lambda(\Phi^{\dag}\Phi)^2$ interaction. We show that at high
temperature and finite density we can use dimensional reduction to
produce an effective three-dimensional theory which then requires
non-perturbative analysis.  For simplicity and ease of
implementation we illustrate this process with the linear delta
expansion.

\end{abstract}

\renewcommand{\thefootnote}{\arabic{footnote}}
\setcounter{footnote}{0}

\section{Introduction}
The inclusion of finite densities of conserved charges in thermal
field theory poses well known problems in the study of phase
transitions. On one hand, perturbative analyses of charged scalar
fields \cite{kap,h&w,ben} give information about the phase
structure, but cannot probe the critical point. Standard
perturbative calculations are plagued with infrared divergences
and after dealing with these, the asymptotic expansion breaks
down. On the other hand, finite charges cannot be easily
represented in lattice Monte-Carlo simulations. Encoding a
non-zero charge in the grand canonical ensemble renders the action
complex and therefore useless as a statistical weighting
\cite{MM}. These obstacles invoke a demand for alternative
methods.

Several authors have recently considered the effect of repulsive
interactions on the condensation temperature of a dilute Bose
gas, a non-relativistic problem. It was pointed out by \cite{mean}
that the leading correction could be isolated in the static
Matsubara mode and a mean field calculation could be performed on
this mode alone. Since then the static mode has been studied using
the $1/N$ expansion \cite{1/N,next1/N}, with the linear delta
expansion \cite{lde}, and now numerically \cite{KPS,AM,AM2}.

In this paper we consider the effect of interactions on the
transition temperature for scalars in the ultra-relativistic
limit. By this we mean that we are at the high temperature limit, but
that the typical self-energy corrections, $\Sigma$, the chemical
potential, $\mu$, or the cube root of the charge density, $\rho$,
can be of the same order $T \gg \Sigma \sim \mu \sim \rho^{1/3}$.
This limit is appropriate for the study of high temperature
symmetry breaking where Bose-Einstein condensation and spontaneous
symmetry breaking have interesting similarities.  The relevant finite
temperature and density 4D Feynman diagrams are difficult to handle
and only Jones and Parkin \cite{phil}
include the setting-sun diagram for the self-energy.
To avoid these problems, we use dimensional reduction
\cite{gin,apple,land,dr-rules} and this is the main focus of our paper.
In our regime, we will show that dimensional reduction for a relativistic
model is much more complicated than in the non-relativistic
limit of our model which studied in
\cite{mean,1/N,next1/N,lde,KPS,AM,AM2}, but we will show it is still
manageable. As in the non-relativistic case, we take advantage of
the good infrared behaviour of the non-static Matsubara modes.
This allows us to integrate these modes perturbatively in an
attempt to simplify our calculation.

Following dimensional reduction, the problem of relativistic
Bose-Einstein condensation has been reduced to the study of a
phase transition in an effective three-dimensional theory at zero
temperature and zero density. Standard non-perturbative methods
can then be employed to study this model but we do not investigate
the many alternatives here, merely choosing one for exemplary
purposes. Here we use the linear delta expansion for its
simplicity and resemblance to standard perturbation theory.

By way of contrast, the only other study of a relativistic finite
density system using dimensional reduction we know of is a study
of QCD \cite{HLP}, rather than the Higgs sector studied here.
The resulting 3D action is complex, making subsequent numerical analysis of \cite{HLP}
more complicated.

\section{Dimensional Reduction}

We begin by considering a relativistic system of bosons described
by a complex scalar field theory. We encode a conserved charge by
working in the grand canonical ensemble with chemical potential
$\mu$. The partition function is then

\be
Z=\int  [d\Psi^{\dag}] [d\Psi] \exp\{-S\}
\ee
where field
integrations are periodic over imaginary time $\beta=1/T$ and the
action is given by
\begin{eqnarray}
S&=&\int_0^{\beta} d\tau\int d^3 x \left\{
    [(\partial_{\tau}+\mu)\Psi^{\dag}][(\partial_{\tau}-\mu)\Psi]
    +\nabla\Psi^{\dag}\nabla\Psi
    +m^2\Psi^{\dag}\Psi
    +\lambda(\Psi^{\dag}\Psi)^2\right\}.
\end{eqnarray}
The charge density is obtained from $Z$ as
follows:
\be
\rho = \frac{T}{V}\frac{1}{Z} \frac{\partial Z}{\partial \mu}
\label{eq:rho}
\ee
where we let the volume $V$ tend to infinity.

The periodicity of the fields is made explicit by a mode expansion
and the non-static modes are integrated
perturbatively\footnote{Alternatively, one can calculate static quantities
in the 3D effective theory and
in the full 4D theory.  By matching these results one can then
relate the coefficients of the 3D
effective theory to those of the 4D one \cite{dr-rules}.  However
there is a lack of results in the literature for our model at
high densities.}. This not only gives an overall factor to $Z$ but
also renormalizes the parameters of the static mode, $\Phi$. The
result is an effective three-dimensional theory whose dependence
on the temperature and chemical potential is contained within its
mass and coupling:

\be
Z=\exp\{ \beta VF(\mu)\}\int [d\Phi^{\dag}]  [d\Phi] \exp\{-S_{\rm 3D}\}.
\label{eq:Z}
\ee

Writing $F=F^{(0)}+\lambda F^{(1)}+{\cal O}(\lambda^2)$, the quadratic part of the action
can be integrated to give
\be
F^{(0)}(\mu)= T\sum_{n\neq 0}\int\frac{d^3 p}{(2\pi)^3}\ln\Delta(\omega_n, p).
\ee
where $\Delta(\omega_n, p)=[(\omega_n-i\mu)^2+p^2+m^2]^{-1}$ and,  since we are dealing
with bosons, $\omega_n = 2\pi n T$. The leading perturbative
correction is given by the figure-of-eight diagram:
\be
F^{(1)}(\mu)=-2T^2\left[\sum_{n\neq 0}\int\frac{d^3 p}{(2\pi)^3}\Delta(\omega_n, p)
    \right]^2.
\ee
We use dimensional
regularization in the $\overline{\rm MS}$ scheme, making the replacement
\be
\int\frac{d^3 p}{(2\pi)^3}\rightarrow\int_{\rm p} =
    \left(\frac{e^{\gamma}M^2}{4\pi}\right)^{\epsilon}
    \int\frac{d^{3-2\epsilon} p}{(2\pi)^{3-2\epsilon}}
\ee
and subtracting only the terms which are divergent as $\epsilon\rightarrow 0$.
$M$ is an arbitrary renormalization scale and $\gamma$ is the Euler-Mascheroni
constant.

The ultra-relativistic limit is defined by $\rho \gg m^3$ or,
equivalently, $T \gg m$. In order to avoid confusion over the
different expansion parameters we will set $m=0$. Since we shall
handle the infrared region non-perturbatively, this does not
create additional problems. We keep corrections of ${\cal
O}(\mu^2/T^2)$ as  we will find these are in fact ${\cal
O}(\lambda)$ for the critical theory. Ignoring any
$\mu$-independent terms, the factor $F$ is then given by (see
\cite{h&w} for details)
\be
F(\mu) = \frac{\mu^2 T^2}{6}\left[1-\frac{\mu^2}{4\pi^2 T^2}
    + \frac{\lambda}{4\pi^2} + {\cal O}\left(\lambda^2\right)\right],
\label{eq:F}
\ee
where by ${\cal O}(\lambda^2)$ we mean ${\cal O}(\mu^4/T^4,\lambda\mu^2/T^2,\lambda^2)$.

\begin{figure}[t]
{\centerline{\resizebox{24pc}{!}{\includegraphics{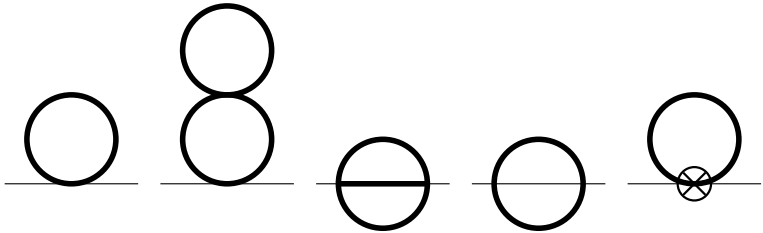}}}}
\caption{Contributions to the mass of the static mode.
    Bold lines correspond to non-static internal lines,
    thin lines correspond to the static mode. The last
    diagram contains the one-loop vertex counterterm.}
\end{figure}

The non-static modes also give corrections to the static mode action. We write the
effective 3D action as
\be
S_{\rm 3D} = \int d^3 x \left[
    \nabla\Phi^{\dag}\nabla\Phi
    +r\Phi^{\dag}\Phi
    +u(\Phi^{\dag}\Phi)^2\right],
\label{eq:S}
\ee
where we neglect the higher dimensional operators $(\Phi^{\dag}\Phi)^n$ for $n\geq3$.
This is firstly on the grounds that their coefficients are high order in $\lambda$
and further that all non-renormalizable interactions are suppressed by factors of
$1/T^2$ (for further discussion of this point see \cite{dr-rules}).
It can be checked that in the calculation which follows, these neglected
terms contribute at higher order in the $\lambda$-expansion than is considered.

The fields should be scaled by $\sqrt{T}$ in order to remove an overall
factor of $1/T$ from the action and to the order we work they undergo
no field renormalization from the non-static mode integration.

Referring to figure 1, the 3D mass is given by
\be
r = -\mu^2+\Sigma_{1}+\Sigma_{2}+\Sigma_{3}+\Sigma_{4}+\Sigma_{5}+{\cal O}(\lambda^3),
\label{eq:m}
\ee
where
\bea
\Sigma_{1}&=&4\lambda T\sum_{n\neq 0}\int_{\rm p}\Delta(\omega_n,p)\\
    &=&\frac{\lambda T^2}{3}\left[1-\frac{3\mu^2}{2\pi^2 T^2}
    + {\cal O}\left(\lambda^2\right)\right];\\
\Sigma_{2}&=&-16\lambda^2 T^2
    \sum_{n_1\neq 0}\int_{\rm p_1}\Delta(\omega_{n_1}, p_1)
    \sum_{n_2\neq 0}\int_{\rm p_2}\Delta^2(\omega_{n_2}, p_2)\\
    &=&-\frac{\lambda^2 T^2}{6\pi^2}\left[\frac{1}{2\epsilon}
    +2\ln\frac{Me^{\gamma}}{4\pi T}+1-\gamma+\frac{\zeta^{/}(-1)}{\zeta(-1)}
    +{\cal O}\left(\lambda\right)\right].
\eea

Though $\Sigma_4$ involves a static internal line it should be
included since it will not arise from the effective static 3D
theory\footnote{Note that at this order, integrating out the
heavy modes actually leaves one with a non-local action with
terms such as $\int d^3x d^3x' |\Phi(x)|^2 B(x,x') | \Phi(x')|^2$ where
$B$ is an $O(\lambda^2)$ bubble diagram. It is when approximating
the 3D theory by the purely local one (\ref{eq:S}) that
contributions, such as $\Sigma_4$, coming purely from non-local
terms in the exact non-local effective theory must not be
forgotten. For instance, the same type of non-local $B$ term
should also lead to a diagram similar to $\Sigma_2$ but with a
light petal on the top and one has to check that this is of lower
order than required. Such problems suggest that the matching of
Green functions approach to dimensional reduction \cite{dr-rules}
might be simpler.}. This contribution is in fact well behaved in
the infrared and to leading order we find (see appendix)
\be
\Sigma_3 +\Sigma_4 = 0+{\cal O}(\lambda^3 T^2).
\ee

Finally we include the diagram with the one-loop vertex counterterm which is given by
$(5\lambda^2 T/8\pi^2\epsilon)(\Phi^{\dag}\Phi)^2$. This is
\bea
\Sigma_{5}&=&4\frac{5\lambda^2 T}{8\pi^2}\frac{1}{\epsilon}
        \sum_{n\neq 0}\int_{\rm p}\Delta(\omega_n,p)\\
    &=&\frac{5\lambda^2 T^2}{12\pi^2}\left[\frac{1}{2\epsilon}
    +\ln\frac{Me^{\gamma}}{4\pi T}+1-\gamma+\frac{\zeta^{/}(-1)}{\zeta(-1)}
    +{\cal O}\left(\lambda\right)\right].
\eea

We may now sum all the contributions to equation (\ref{eq:m}) giving
\be
r=-\mu^2
    +\frac{\lambda T^2}{3}\left[1-\frac{3\mu^2}{2\pi^2T^2}
    +\frac{3\lambda}{4\pi^2}\left(\frac{1}{2\epsilon}+\frac{1}{3}
    \ln\frac{Me^{\gamma}}{4\pi T}+1-\gamma
    +\frac{\zeta^{/}(-1)}{\zeta(-1)}\right)
    +{\cal O}\left(\lambda^2 \right)\right].
\label{eq:m2}
\ee

We shall find the coupling to be given with sufficient accuracy by
\be
u = \lambda T+{\cal O}(\lambda^2).
\label{eq:lambda}
\ee
The coupling is now dimensionful due to the scaling of the static fields
by $\sqrt{T}$.

Use of equations (\ref{eq:rho}) and (\ref{eq:Z}) with (\ref{eq:F}), (\ref{eq:S}),
(\ref{eq:m2}) and  (\ref{eq:lambda}) gives
\bea
\rho &=& \frac{\partial F}{\partial\mu}+2\mu T
    \left[1+\frac{\lambda}{2\pi}+{\cal O}\left(\lambda^2\right)
    \right]\langle\Phi^{\dag}\Phi\rangle\\
    &=& \frac{\mu T^2}{3}\left[1-\frac{\mu^2}{2\pi^2 T^2}
    +\frac{\lambda}{4\pi^2}+{\cal O}\left(\lambda^2\right)\right]
    +2\mu T\left[1+\frac{\lambda}{2\pi^2}
    +{\cal O}\left(\lambda^2\right)\right]
    \langle\Phi^{\dag}\Phi\rangle,
    \label{rhoexp1}
\eea
where $\langle\Phi^{\dag}\Phi\rangle$ denotes
the Green function evaluated in the
effective 3D theory. This we cannot calculate
perturbatively since the expansion
will break down when probing large length scales greater than $1/u$.

\section{Linear Delta Expansion}

To evaluate the quantity $\langle\Phi^{\dag}\Phi\rangle$ we need a
non-perturbative method as the 3D sector retains all the infrared
divergences of the 4D theory.  The effective 3D theory is studied
at zero temperature and density so the problem is greatly
simplified.  Any standard non-perturbative method can be used at
this point.

We will use LDE, the Linear Delta Expansion, though the method
is also known by several other names (see \cite{EJW} for a brief
summary). LDE has been used successfully in many situations,
including studies of scalar theories at non-zero density such as
\cite{lde,EJR,EIM,phil,EJW}.  In toy models, where exact results
are achievable, LDE is known to produce convergent results and to do
so much faster than alternatives, for instance see \cite{DJ,BeDJ}
and references therein.  In full QFT, LDE has also often proved to
be better than other methods \cite{PO}.

We begin by defining $S_\delta$ which interpolates between our 3D
action (from which we now drop the subscript) and some soluble
action $S_0$ as $\delta$ varies from 1 to 0:

\be
S \rightarrow S_{\delta} = \delta S + (1-\delta) S_0.
\ee
We are free to choose
\be
S_0 = \int d^3 x \left[\nabla\Phi^{\dag}\nabla\Phi
    +\Omega^2\Phi^{\dag}\Phi\right]
\ee
such that
\be
S_{\delta}=\int d^3 x \left[\nabla\Phi^{\dag}\nabla\Phi
    +(\Omega^2-\delta\Omega^2+\delta r)\Phi^{\dag}\Phi
    +\delta u(\Phi^{\dag}\Phi)^2 \right].
\ee

Any physical quantity ${\cal P}$ is evaluated as a power series
in $\delta$ to some finite order. This quantity will generally have some
dependence on $\Omega$ which we fix by some specified criterion which we take to
be the principal of minimal sensitivity (PMS) \cite{stevenson}:
\be
\left.\frac{d{\cal P}}{d\Omega}\right|_{\delta=1,\Omega={\overline \Omega}}=0 .
\ee
This variational procedure allows for the emergence of non-perturbative
behaviour.

The Green function we require can now be written as
\be
\langle\Phi^{\dag}\Phi\rangle = \int\frac{d^3 p}{(2\pi)^3} G_{\delta}(p),
\label{eq:green}
\ee
where
\be
G_{\delta}(p) = \left[p^2 + \Omega^2 - \delta\Omega^2 +\delta r
    +\Sigma_{\delta}(p)\right]^{-1}.
\ee

\begin{figure}[t]
{\centerline{\resizebox{20pc}{!}{\includegraphics{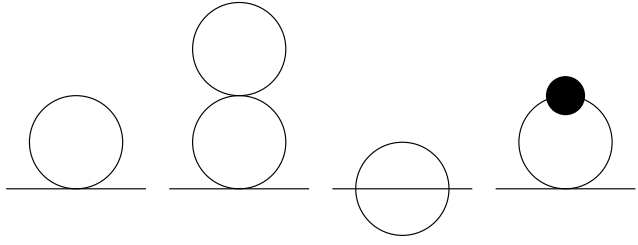}}}}
\caption{Contributions to the self-energy for the effective 3D
theory. The filled
    dot corresponds to a $\delta(r-\Omega^2)$ mass insertion.}
\end{figure}

The criterion defining the transition temperature is that the correlation length
in the original theory be infinite. This can be expressed as
$\left.G^{-1}_{\delta}(0)\right|_{\delta=1}=0$ which is satisfied by imposing
\be
\delta r +\Sigma_{\delta}(0)=0.
\label{eq:crit}
\ee
Use of this relation in equation (\ref{eq:green}) gives
\be
\langle\Phi^{\dag}\Phi\rangle = \int\frac{d^3 p}{(2\pi)^3}
    \left[p^2 + \Omega^2 - \delta\Omega^2 + \Sigma_{\delta}(p)
    - \Sigma_{\delta}(0)\right]^{-1}
\ee
and expanding to second order in $\delta$ we have
\be
\langle\Phi^{\dag}\Phi\rangle = \int\frac{d^3 p}{(2\pi)^3}
    \frac{1}{p^2+\Omega^2}\left[1+\frac{\delta\Omega^2}{p^2+\Omega^2}
    +\frac{\delta^2\Omega^4}{(p^2+\Omega^2)^2}
    -\frac{\Sigma_{\delta}(p)-\Sigma_{\delta}(0)}{p^2+\Omega^2}\right]
    +{\cal O}(\delta^3).
\ee
The only momentum dependent contribution to the self energy results from the
setting sun diagram (see figure 2):
\be
\Sigma^{\rm 3D}_{3}(p) = -8\delta^2 u^2\int\frac{d^3 k}{(2\pi)^3}
    \frac{d^3 q}{(2\pi)^3}
    \frac{1}{(k^2+\Omega^2)}\frac{1}{(q^2+\Omega^2)}\frac{1}{((k+q+p)^2+\Omega^2)}.
\label{eq:sun}
\ee
We continue our use of dimensional regularization in the $\overline{\rm MS}$ scheme, using the same scale $M$ as in our 4D heavy mode calculations.
The required integrals (see, for example, \cite{lde} and \cite{int}) are
\bea
\int_{\rm p}\frac{1}{p^2+\Omega^2}&=&-\frac{\Omega}{4\pi}\left[1+2\epsilon
    \left(\ln\frac{M}{2\Omega}+1\right)+{\cal O}(\epsilon^2)\right],\label{eq:int}\\
\Sigma^{\rm 3D}_{3}(0)&=&\frac{8\delta^2 u^2}{(4\pi)^2}\left[
    \frac{1}{4\epsilon}+\ln\frac{M}{3\Omega}+\frac{1}{2}
    +{\cal O}(\epsilon)\right],\\
\int_{\rm p}\frac{1}{p^2+\Omega^2}\Sigma^{\rm 3D}_{3}(p)&=&-\frac{2\delta^2 u^2}
    {(4\pi)^3\Omega}\left[\frac{1}{2\epsilon}+3\ln\frac{M}{2\Omega}+1
    -2\ln2+{\cal O}(\epsilon)\right].
\eea
Further integrals may be derived from (\ref{eq:int}) by successive
differentiations with respect to $\Omega$.

Summing the contributions to the Green function, we find
\be
\langle\Phi^{\dag}\Phi\rangle =-\frac{\Omega}{4\pi}+\frac{\delta\Omega}{2(4\pi)}
    +\frac{\delta^2\Omega}{8(4\pi)}-\frac{4\delta^2 u^2}{(4\pi)^3\Omega}
    \ln\frac{4}{3}.
\ee
The divergences cancel and there is no need to invoke any counter
terms. We now apply the PMS condition to $\langle
\Phi^{\dag}\Phi\rangle$ and arrive at
\be
{\overline \Omega}=\pm \frac{u}{\pi}\left(\frac{2}{3}\ln\frac{4}{3}\right)^\frac{1}{2}.
\ee
The question of which solution to choose can be answered by comparison with the
solution in the large-$N$ case for $N\rightarrow 2$. Referring to
\cite{1/N} and
\cite{next1/N} we find that the positive solution is appropriate, giving
\bea
\langle\Phi^{\dag}\Phi\rangle &=& -\frac{u \bar{f} }{(4\pi)^2} ,
 \label{fbardef}
 \\
 \bar{f} &=& \left(6\ln\frac{4}{3}
    \right)^{\frac{1}{2}} \approx 1.314
    .
    \label{fbarbeval}
\eea
Inserting this into equation (\ref{rhoexp1}) we have for the
critical density
\be
\rho = \frac{\mu T^2}{3}
  \left[ 1
         - \frac{\mu^2}{2\pi^2 T^2}
         + \frac{\lambda}{(4\pi)^2}
         \left( 4-6\bar{f} \right)
         + {\cal O}\left(\lambda^2\right)
  \right] .
 \label{rhocrit1}
\ee

The chemical potential is an unwanted free variable in this expression and we must
use relation (\ref{eq:crit}) to constrain $\mu$ at the
transition temperature. Along with the setting sun contribution to the self energy,
we also have to order $\delta^2$ (see figure 2)
\bea
\Sigma^{\rm 3D}_{1}&=&4\delta u\int_{\rm p}\frac{1}{p^2+\Omega^2},\\
\Sigma^{\rm 3D}_{2}&=&-16\delta^2 u^2\int_{\rm p}\frac{1}{(p^2+\Omega^2)^2}
    \int_{\rm q}\frac{1}{q^2+\Omega^2},\\
\Sigma^{\rm 3D}_{4}&=&4\delta^2 u (\Omega^2-r)\int_{\rm p}
    \frac{1}{(p^2+\Omega^2)^2}.
\eea
Evaluating and summing these contributions gives
\be
\Sigma_{\delta}(p=0)=-\frac{\delta u \Omega}{\pi}
    -\frac{8\delta^2 u^2}{(4\pi)^2}\left[\frac{1}{4\epsilon}+
    \ln\frac{M}{3\Omega}-\frac{1}{2}\right]+\frac{\delta^2 u \Omega}{2\pi}
    -\frac{\delta^2 u r}{2\pi \Omega}+ {\cal O}(\delta^3),
\ee
which upon substituting into (\ref{eq:crit}) gives
\be
 \delta
 r=\frac{\delta u \Omega}{\pi}-\frac{\delta^2 u \Omega}{2\pi}
    +\frac{\delta^2 u^2}{2\pi^2}\left[\frac{1}{4\epsilon}+
    \ln\frac{M}{3\Omega}+\frac{1}{2}\right]+ {\cal O}(\delta^3).
\label{eq:r}
\ee
We apply the PMS condition to $r$, giving ${\overline \Omega} =
u/\pi$, and insert this value into equation (\ref{eq:r}):
\bea
r &=& \frac{u^2}{2\pi^2} \left[ \frac{1}{4\epsilon} +
 \ln\frac{M}{u} + c_r \right]
 \label{crdef},
 \\
 c_r &=& \ln\frac{\pi}{3} + \frac{3}{2}  \approx 1.546 .
 \label{crbeval}
 \eea
The dependence of $r$ on the scale $M$ is exact because of
superrenormalisability and agrees with that noted elsewhere (e.g.\
\cite{AM2}) even
though it appears here in the context of a particular non-perturbative
calculation.
Comparing with equation (\ref{eq:m2}) we have
\bea
-\mu^2+\frac{\lambda T^2}{3}\left[1-\frac{3\mu^2}{2\pi^2T^2}
    +\frac{3\lambda}{4\pi^2}\left(\frac{1}{2\epsilon}+\frac{1}{3}
    \ln\frac{Me^{\gamma}}{4\pi T}+1-\gamma
    +\frac{\zeta^{/}(-1)}{\zeta(-1)}\right)+{\cal O}(\lambda^2)
    \right]\nonumber\\
    =\frac{\lambda^2 T^2}{2\pi^2}\left[\frac{1}{4\epsilon}
    +\ln\frac{M}{\lambda T} + c_r\right].
\eea
The divergences cancel and we can obtain $\mu$ in terms of the critical temperature:
\bea
\mu &=&
 \frac{\sqrt{\lambda}T}{\sqrt{3}}\left[1+\frac{\lambda}{8\pi^2}
  \left( \alpha_{\rm mu} \ln\frac{T}{M}+ \eta_{\rm mu} \ln\lambda +  a_{\rm mu} \right)
    +{\cal O}(\lambda^2)
    \right],
\label{mures}
 \\
 && \alpha_{\rm mu} = 5, \; \; \; \eta_{\rm mu} = 6,
 \\ &&
 a_{\rm mu} = \ln \left( \frac{e^\gamma}{4 \pi} \right)
 + 1 +3 \left( -\gamma + \frac{\zeta(1,-1) }{\zeta(-1)}   \right)
    -6 c_r
 \approx 3.270 - 6 c_r \approx -6.007 .
\label{muberes}
\eea
It is now clear that for the critical theory, $\mu^2/T^2\sim{\cal
O}(\lambda)$. Substituting into the equation for the charge
density gives
\bea
\rho &=&
\frac{\sqrt{\lambda}T^3}{3\sqrt{3}}\left[1+\frac{\lambda}{8\pi^2}
    \left(\alpha_{\mathrm rho} \ln\left(\frac{T}{M}\right)
    + \eta_{\mathrm rho} \ln (\lambda)
    + a_{\mathrm rho}
    \right)+{\cal O}(\lambda^2)\right],
 \label{eq:result}
 \\  &&
  \alpha_{\mathrm rho} = 5, \;\;\;
  \eta_{\mathrm rho} = 6,
  \\ &&
  a_{\mathrm rho} = a_{\rm mu} +\frac{2}{3} - 3 \bar{f}
  \approx  3.9236 - 6c_r - 3 \bar{f} \approx - 21.759 .
  \label{rhoberes}
\eea
Equation (\ref{eq:result}) relates the critical density to the
critical temperature for the Bose-Einstein condensation of a
complex scalar field in the ultra-relativistic limit. Though
(\ref{eq:result}) looks like an expansion in $\lambda$ we stress
that the result is non-perturbative because of the severe IR
problems in calculating $\langle\Phi^{\dag}\Phi\rangle$ and
self-energies at the critical point in the three-dimensional
theory. We should expect $\rho$ to have some renormalization scale
dependence since we have included a one-loop vertex counterterm in
the calculation.

\section{Discussion}

The leading term in our expressions
for the critical chemical potential (\ref{mures}) or equivalently
for the critical density (\ref{eq:result}), are the usual leading
high temperature results, $\mu_0 = \sqrt{\lambda /3}T$ and $\rho_0
= \sqrt{\lambda/27}T^3$.

The $\lambda^2 \ln(T/M)$ term is
{\em exactly} that expected from the running of the 4D coupling
$\lambda$ in the leading term using perturbation theory where one finds
\begin{equation}
\lambda(M_2) = \lambda(M_1)
+ \frac{5 \lambda^2}{8 \pi^2} \ln \left( \frac{M^2_2}{M_1^2} \right) .
\end{equation}
The perturbative result is appropriate as the leading behaviour comes only
from the heavy modes, and these are dealt with perturbatively in this
calculation.  Thus we find that once the implicit scale dependence of
$\lambda$ is accounted for, our results for $\mu$ and $\rho$ are actually
independent of the scale $M$ as all exact physics results should be.

The $\lambda^2 \ln (\lambda)$ term comes directly from our
expression (\ref{crdef}) for $r$ and in particular comes from dependence
on the scale $M$ which is exact for for the super renormalisable theory.
In the context of a Green
function matching approach, as discussed for instance in
\cite{dr-rules}, this term might be described as running the 3D mass
$r$ from the scale $M \sim T$, used when dealing with the heavy modes,
down to $u = \lambda T$ appropriate for the 3D static theory.
Overall then our expressions for the critical
chemical potential and density
agree with our expectations from other calculations.

In terms of actual numbers, for $\lambda = 1/8$ the fractional
corrections to this leading term are of the order of a few
percent, e.g.\ in units of the scale $M$, for $T=10.0$, the
critical chemical potential is $\mu=\mu_0 \times (1-0.010)$,
$\mu_0=2.04$ and the critical density is $\rho = \rho_0 \times (1 -
0.058)$, $\rho_0=68.0$.

A good test of our central result, the dimensional reduction,
is to compare our formulae for the critical chemical potential
with that extracted from the results of Jones and
Parkin \cite{phil}. They also use the linear delta expansion but they
apply it directly to the full four-dimensional theory
anywhere in the symmetric
phase.   They can in principle study any temperature and density,
though they limit their analysis to high temperatures. We
use dimensional reduction so our method is always limited to high
temperatures.  However, this brings several benefits to us as
further fields, including fermions,
can be added with great ease and our non-perturbative
effort is much less. The effective three-dimensional theory
derived here can in principle be studied using any
non-perturbative technique, including Monte-Carlo since
the effective action
is real. While we have only looked at the critical point, we have
also given the critical density which is the actual measured
quantity.  Finally, our results are completely analytic, while
Jones and Parkin can only find numerical solutions to their
equations, though the numerics are relatively straight forward.

\begin{figure}[t]
{\centerline{\resizebox{24pc}{!}{\includegraphics{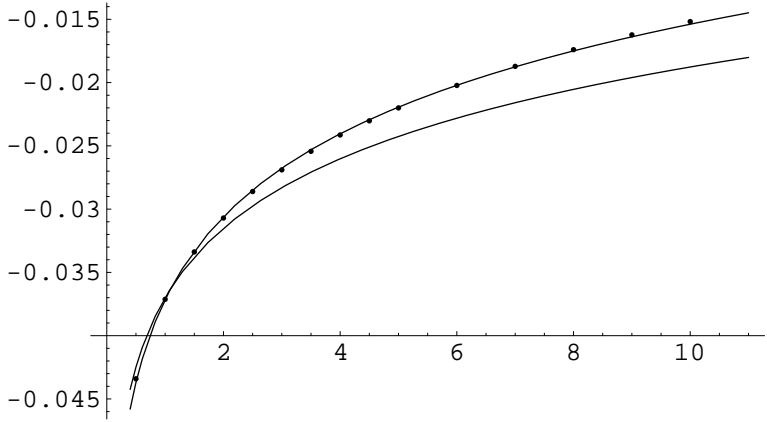}}}}
\caption{Fractional $\mu$ correction, $\mu/\mu_0 -1$, against the
critical $T$ in units of $M$ for $\lambda = 1/8$. Dots are data
points from the full 4D Jones and Parkin method with best fit
curve connecting them. Other line is the dimensional reduction
form with LDE used for 3D non-perturbative
calculation.}\label{fjpvsbe}
\end{figure}

Turning to the details of the results, we find that the
Jones-Parkin method gives the same qualitative behaviour as our
results for $\lambda = 1/8$, $T/M=0.5 \ldots 10.0$.  The
fractional correction to the leading $\mu$ behaviour, $\Delta \mu
:= \mu/\mu_0 -1$ is shown in figure \ref{fjpvsbe}. However their
results for these parameters \cite{JPunpub} are best fitted with
slightly different coefficients\footnote{A shift from MS to
$\overline{\rm MS}$ scales is needed.}, namely $\alpha_{\mathrm
rho,JP} / \alpha_{\mathrm rho} = 1.2(1)$, and choosing
$\eta_{\mathrm rho,JP} = \eta_{\mathrm rho}$ we find $a_{\mathrm
rho,JP} + \eta_{\mathrm rho,JP} \ln \lambda= -17.8(2)$ compared
with our value $-18.5$. In our calculation the $\alpha_{\mathrm
mu}$ value was set by perturbation theory and can not be exact.
Since Jones and Parkin use an entirely non-perturbative method,
even though both methods ought to be valid in the parameter range
considered, there are likely to be small differences between the
two results. These ought to be $O(\lambda)$ fractional corrections
so the results seem to be consistent as far as they go.

Finally, we can illustrate the power of our work by noting that
any non-perturbative calculation can be used with our
formulae\footnote{Note that the normalisation of fields and the definition of
$u$ or $\lambda$ often differ from those used here by simple constant
factors.} for dimensional reduction in the presence of a
large density. For instance, since this work was completed, two
lattice Monte Carlo studies of the 3D action relevant here have
appeared \cite{KPS,AM,AM2}.  While these numerical results
were produced for the non-relativistic
4D Bose-Einstein condensation problem, we can just as easily apply
them to our relativistic case.
The two studies are completely consistent but for definiteness we
use the values in \cite{AM2} which gives $c_{\mathrm
r,lat} = 0.671 (1)$ and $\bar{f}_{\mathrm lat} = 0.57 (1) $.  In
this case, the $\alpha$ and $\eta$ terms of both $\mu$ and $\rho$
expressions are fixed by dimensional reduction so the only
difference is in the $a$ parameters, and we find that the lattice
data leads to $a_{\mathrm mu, lat}=-0.756$ and $a_{\mathrm rho,
lat}=-13.2$ compared to our values of $-6.01$  and $-21.8$
respectively. While these appear to be large differences, in
physical quantities such as the critical density $\rho$
the constant coming from $a$ is in fact overwhelmed by the contribution
from the $\ln \lambda$ term for $\lambda \ll 1 $ which is where the
dimensional reduction process is valid.  The difference is more of
a comment on the efficacy of different non-perturbative
approximations in 3D calculations (see \cite{AM} for a good
comparison) than particularly important to our results.

In conclusion, we have shown that by organising the modes into
those which can be pertubatively integrated and those which
cannot, we minimise the non-perturbative effort needed
to study Bose-Einstein condensation
at relativistic temperatures and densities. The method
is economical and provides a reliable estimate of the critical
temperature and critical density.


\section*{Acknowledgements}
We would like to thank H.F.Jones and P.Parkin for providing us
with some of their data.  DJB was supported by PPARC.



\renewcommand{\theequation}{A.\arabic{equation}}
\setcounter{equation}{0}

\section*{Appendix}
In this appendix we consider how the sunset diagram can be decomposed
into contributions from different modes.

We begin by considering the $\mu=0$ case and state the results
(see, for example, \cite{3dsum,arnold})
\bea
\sum_{n_1,n_2=-\infty}^{\infty}\int_{\rm {p_1,p_2}}
        \frac{1}{(\omega_{n_1}^2+p_1^2+m^2)}\frac{1}
    {(\omega_{n_2}^2+p_2^2+m^2)}\frac{1}{((\omega_{n_1}+\omega_{n_2})^2
    +(p_1+p_2)^2+m^2)}\nonumber\\
=\frac{1}{(4\pi)^2}\left[
        \frac{1}{4\epsilon}+\ln\frac{M}{3m}+\frac{1}{2}\right]
        +{\cal O}\left(\frac{m}{T}\right)
\label{eq:sunny}
\eea
and
\be
\int_{\rm {p_1,p_2}}\frac{1}{(p_1^2+m^2)}\frac{1}{(p_2^2+m^2)}
\frac{1}{((p_1+p_2)^2+m^2)}=\frac{1}{(4\pi)^2}\left[
        \frac{1}{4\epsilon}+\ln\frac{M}{3m}+\frac{1}{2}\right].
\label{eq:sunny3d}
\ee
We shall split (\ref{eq:sunny}) into purely non-static internal lines
($n_1\neq 0; n_2\neq 0; n_1\neq n_2$),
one static internal line,
($n_1\neq 0; n_2\neq 0; n_1=n_2$ and $n_1\neq 0; n_2=0$ and $n_1=0;
n_2\neq 0$), and purely static internal lines, ($n_1=0; n_2=0$).
This covers all the possibilities and we may write
\be
\sum_{n_1,n_2=-\infty}^{\infty}=
    \sum_{\stackrel{n_1,n_2\neq 0}{n_1\neq n_2}}\left.\right.
    +6 \delta_{n_1}\sum_{n_2>0}\left.\right.
    +\delta_{n_1}\delta_{n_2}
\ee
Denoting these contributions to (\ref{eq:sunny})
as $I_{\rm nonstatic}$, $I_{\rm mixed}$
and $I_{\rm static}$, we immediately
see that $I_{\rm static}$ is given by equation (\ref{eq:sunny3d}).

The contribution with one static line can be calculated by taking the
static line to be massless. This does not cause infrared divergences and
is appropriate for the critical theory. We may also set $m=0$ in the other
propagators since $T\gg m$. We thus have
\bea
I_{\rm mixed}&=&
    6\sum_{n>0}\int_{\rm p_1,p_2} \frac{1}{p_1^2}\frac{1}{p_2^2+\omega_n^2}
    \frac{1}{(p_1+p_2)^2+\omega_n^2}+{\cal O}\left(\frac{m}{T}\right)\\
&=&6\sum_{n>0}\left(\frac{e^{\gamma}M^2}{4\pi}\right)^{\epsilon}
    \frac{\Gamma^3(1/2+\epsilon)}{(4\pi)^{3/2-\epsilon}\Gamma(1+2\epsilon)}
    \int_{\rm p_1}\frac{1}{p_1^2}\frac{1}
    {(p_1^2+\omega_n^2)^{1/2+\epsilon}}+{\cal O}\left(\frac{m}{T}\right)\\
&=&6\sum_{n>0}-\left(\frac{e^{\gamma}M^2}{4\pi \omega_n^2}\right)^{2\epsilon}
    \frac{\Gamma^2(1/2+\epsilon)}
    {2\epsilon(\epsilon-1/2) (4\pi)^{3-2\epsilon}}
    +{\cal O}\left(\frac{m}{T}\right)\\
&=&-\left(\frac{e^{\gamma}M^2}{16\pi^3 T^2}\right)^{2\epsilon}
    \frac{3\zeta(4\epsilon)\Gamma^2(1/2+\epsilon)}
    {\epsilon(\epsilon-1/2)(4\pi)^{3-2\epsilon}}
    +{\cal O}\left(\frac{m}{T}\right)\\
&=&-\frac{3}{(4\pi)^2}\left[\frac{1}{4\epsilon}+\ln{\frac{M}{2T}}
    +\frac{1}{2}\right]+{\cal O}\left(\frac{m}{T}\right).
\eea

Finally, the purely non-static contribution can be found
 by subtracting the other contributions from (\ref{eq:sunny}):
\be
I_{\rm nonstatic}=
\frac{3}{(4\pi)^2}\left[\frac{1}{4\epsilon}+\ln{\frac{M}{2T}}
    +\frac{1}{2}\right] +{\cal O}\left(\frac{m}{T}\right).
\ee

Turning to the case where $\mu\neq 0$, $I_{\rm nonstatic}$ is unchanged at
leading order in the high temperature expansion since
$\mu\ll T$ and we may clearly take the $\mu\rightarrow 0$ limit without
causing infrared divergences. $I_{\rm mixed}$ is also unchanged upon
choosing the static line to be critical. Up to corrections of
${\cal O}(\mu^2/T^2)$ and ${\cal O}(m/T)$, the non-static and mixed
contributions to the sunset diagram sum to zero.

\end{document}